\documentclass[aps,prl,twocolumn,noshowpacs,superscriptaddress,groupedaddress]{revtex4}  
\usepackage{graphicx}  
\usepackage{dcolumn}   
\usepackage{bm}        
\usepackage{amssymb}   

\usepackage[T2A]{fontenc}

\usepackage[english,russian]{babel}

\usepackage{amsmath}

\usepackage{amsfonts}

\usepackage{textcomp}

\usepackage{bm}

\usepackage{color}

\usepackage{verbatim}

\usepackage{subfig}

\begin{document}

\title{Notes on conservation laws in chiral hydrodynamics}


\author{V.I. Zakharov}
\affiliation{ITEP, B. Cheremushkinskaya 25, Moscow, 117218 Russia}
\affiliation{Moscow Inst Phys \& Technol, Dolgoprudny, Moscow Region, 141700 Russia.}
\affiliation{Far-East Federal University, Vladivostok, School of Medicine}

\begin{abstract}
We  consider chiral fluids 
within the standard framework of a chiral-invariant underlying field theory,
anomalous in presence of electromagnetic fields. 
Apart from the Noether axial current of the underlying theory,
in the limit of ideal fluid there exist extra conserved currents, corresponding 
to classical helical motions. 
The extra conservation laws are known to break down once 
viscosity is non-vanishing. Which looks puzzling, as if introduction of viscosity were 
inconsistent with chiral invariance. As a resolution of the puzzle, we  
 argue
that locally one can introduce an inertial frame where an extra conservation law
still holds.  
 In other words, the extra currents are covariantly conserved.
The emergent  gravitational field is determined by  dynamics of the viscous  fluid.
We turn  then to instabilities of chiral plasma against decays into  helical
magnetic or vortical
configurations. We emphasise similarity between the two cases in the far infrared
region, responsible for the decays.
This similarity is not apparent within the standard counting of 
orders in derivative expansion.
The material was originally prepared for a review talk by the author.

\end{abstract}

\maketitle

\section{Introduction.}

Chiral media
attracted a lot of attention recently, see, in particular, 
the collection of review articles
\cite{volume} and references therein. 
There are a few remarkable effects
to be observed in such media. First of all, the chiral anomaly 
\cite{adler} is  manifested
macroscopically,  already in the hydrodynamic approximation 
\cite{cheianov,sonsurowka,oz}. In more
detail, one starts with a fundamental theory possessing an 
anomalous $U(1)$ chiral symmetry. The
 fundamental conservation laws read as:
\begin{eqnarray}\label{conservationlaws}
\partial_{\mu}\Theta^{\mu\nu}~=~eF^{\nu\rho}j_{\rho}~,\\
\partial_{\sigma}j^{\sigma}_{el}~=~0~,~~~\partial_{\sigma}j^{\sigma}_5~=~
e^2C\tilde{F}^{\mu\nu}F_{\mu\nu} \nonumber
\end{eqnarray}
where $\Theta^{\mu\nu}, j^{\mu}_{el}, j^{\mu}_5$ are the energy-momentum tensor, 
electromagnetic current (with $e$ being the corresponding coupling)  
and axial-vector current, respectively, and 
$F_{\mu\nu}$ is the electromagnetic field-strength
tensor, $\tilde{F}^{\mu\nu}~=~
1/2\epsilon^{\mu\nu\rho\sigma}F_{\rho\sigma}$ and $C$ is the
coefficient in front of the chiral anomaly, 
$C=1/(12\pi^2)$ for one flavour doublet. 

The hydrodynamic equations of motion follow from (\ref{conservationlaws})
provided that $\Theta^{\mu\nu},j^{\mu}_{el}, j^{\mu}_5$
are put into constituent form.
Furthermore, Eqs
(\ref{conservationlaws}) are supplemented
by  an equation of state. In most cases, non-vanishing 
 chemical potentials, $\mu, \mu^5$  and 
finite temperature $T$ are introduced at this point. For our purposes,
 we do not need 
to specify the equation of state. Moreover, 
to simplify presentation, we will
mostly omit  temperature-dependent terms.
 
Remarkably enough, using only such a general framework one can  fix  
\cite{erdmenger,sonsurowka,oz} the 
so called chiral magnetic and chiral vortical effects.
By the chiral magnetic effect \cite{vilenkin,cheianov,kharzeev1}
one understands
a flow of electric current along an external magnetic field. In case of 
an electrically neutral but chirally asymmetric plasma one gets:
\begin{equation}\label{chme}
j^{\rho}_{el}~=~Ce\mu_5 B^{\rho}~,
\end{equation} 
where $\mu_5$ is the chiral chemical potential, 
$B^{\rho}~=~(1/2)\epsilon^{\rho\sigma\alpha\beta}u_{\sigma}F_{\alpha\beta}$, 
$u^{\sigma}$ is the 4-velocity of an 
element of the liquid, and the constant $C$
is anomaly-related, as defined above. 
The chiral vortical effect \cite{vilenkin2,erdmenger,sonsurowka}, 
in turn, refers to the axial-vector current 
proportional to
the vorticity $\omega^{\alpha}$ of the liquid, 
$\omega^{\alpha}~=1/2\epsilon^{\alpha\beta\gamma\delta}
u_{\beta}\partial_{\gamma}u_{\delta} $.
If both $\mu$ and $\mu_5$ are nonvanishing the axial curent is given by:
\begin{equation}\label{chve}
j^{\rho}_5~=~n_5u^{\rho}+C(\mu^2+\mu_5^2)\omega^{\rho}~+~O(e),
\end{equation}
where $n_5~\equiv~1/2(n_R-n_L)$ is the the density of the chiral charge,
and $O(e)$ are terms vanishing in the limit of the electromagnetic coupling tending 
to zero.
 
Solutions  (\ref{chme}), (\ref{chve}) exhibit a few interesting features.
In particular, one can switch off
electromagnetic  fields so that the axial current is non-anomalous.
Nevertheless, the chiral vortical effect (see (\ref{chve})) survives and is proportional to
the coefficient $C$ which apparently encodes the anomaly. 
Moreover, since the naive axial current
is conserved in this limit, Eq. (\ref{chve}) suggests 
that the chiral vortical current, or the second term in 
the r.h.s. of Eq. (\ref{chve}) is separately conserved as well.

And, indeed, it is known since long, see, e.g. \cite{bekenstein},
that in the non-dissipative limit fluids possess extra conservation
laws. 
In other words, hydrodynamics of ideal fluids possesses higher symmetry than
the fundamental
field theory behind it. 
In view of this, one could argue that the ideal-fluid approximation might be 
a wrong starting point to study the general case of dissipative fluids. However,
the ideal fluid approximation is singled out 
since it is the leading term in the hydrodynamic expansion (in derivatives).
Also, for ideal liquids one can develop a fully  field-theoretic description
in the sense that infrared degrees of freedom can be identified in all
the generality and reduce to a set of scalar fields, see, e.g., \cite{nicolis,abanov} 
and references therein. To include dissipation, one needs to introduce in field-theoretic language
violation of unitarity and relation of hydrodynamics to
fundamental field theory is much less transparent. 

This extra symmetry is broken once viscosity is taken into account
\cite{bekenstein}. Indeed,  
upon inclusion of effects of viscosity $\eta \neq 0$, the motion would
 end up
in most cases with a trivial equilibrium configuration of the whole fluid being at rest
(for exceptional cases see \cite{bhattacharaya}).
Thus, it seems rather  obvious that the fluid helicity is not conserved for viscous fluids.
It is not  straightforward at all, however,
to appreciate this observation in field-theoretic terms. 
 For example, a recipe
 to construct an anomalous axial current in non-equilibrium
is suggested in Ref. \cite{rangamani}.
The current is formulated in terms of the Keldysh-Schwinger contour
and is highly non-local.

In these notes we try to bridge the cases with $\eta \equiv 0$ and
$\eta \neq 0$, but small. The basic idea is that in case of chiral fluids
inclusion of effects of viscosity intrinsically, or dynamically does not trigger
violation of chirality conservation. There is, however, 
a kinematical effect which can be described in terms
of locally inertial frames.
In more detail, Eqs. (\ref{chme}), (\ref{chve})
assume a universal, Lorentz-covariant description
of the whole fluid in terms of the 4-vector $u_{\mu}(x)$.
Inclusion  of viscosity introduces acceleration $a^{\mu}=du^{\mu}/ds$
which depends on position $x^{\mu}$ and is determined by dynamics of the
fluid (say, through the Navier-Stokes equations).
Under these circumstances, the conservation laws would 
preserve their form in a locally inertial frame while
in other frames the conservation laws would look as vanishing of
the covariant derivative of the corresponding current.

In particular, introduce the fluid-helicity current \cite{bekenstein}
defined as:
\begin{equation}\label{fluid-hel}
j^{\alpha}_{fluid~hel}~\equiv~\epsilon^{\alpha\beta\gamma\delta}
(\mu\cdot u_{\beta})\partial_{\gamma}(\mu\cdot u_{\delta})~~,
\end{equation}
where $\mu$ is the chemical potential
(and $\mu_5=0$, for simplicity).
In case of ideal fluid the current (\ref{fluid-hel}) is conserved, 
for a recent discussion and further references see, e.g., \cite{avdoshkin}:
\begin{equation}
\partial_{\alpha}j^{\alpha}_{fluid~hel}~=~0~~~~(ideal~fluid)~~.
\end{equation}
We suggest that (at least for a small viscosity), the current 
(\ref{fluid-hel}) is covariantly conserved for  viscous fluid:
\begin{equation}\label{crucial}
D^{eff}_{\alpha}j^{\alpha}_{fluid~hel}~=~0~~~~~(viscous~ fluid),
\end{equation}
where the ``effective'' covariant derivative
$D^{eff}_{\alpha}$  is defined algebraically the same
as in general relativity but the  standard external gravitational field
entering the covariant derivative
is replaced by an
effective gravitational field   $h^{eff}_{\mu\nu}$ 
determined by the fluid dynamics and, for small viscosity $\eta$
the effective field $h^{eff}_{\mu\nu}~\sim~ \eta$.
The physical
meaning of (\ref{crucial}) is transparent.
Namely, in a chiral invariant theory viscosity does not introduce 
any dynamical violation of conservation of the axial current.
In the viscous case, however,
it is not possible any longer to introduce a global inertial frame
where Eq. (\ref{chve}) would  hold. 

We do check (\ref{crucial}) in a trivial case when the 
physical effect is linear in $h^{eff}_{\mu\nu}$ (and $\eta$). As for the higher
order terms, validity of (\ref{crucial}) 
remains a hypothesis \footnote{Note, though, that, in other contexts, 
a somewhat similar problem of finding a mapping
of solutions of the Navier-Stokes equations into solutions of
Einstein equations has been discussed in great detail
and the mapping explicitly constructed,
 see, in particular \cite{skenderis} and references therein}.
 
The organisation of the paper is as follows. In Sect. 2 we discuss conservation
of various types of helicity in case of ideal fluid. In Sect. 3
we present in more detail motivation to introduce emergent gravitational field.
In Sect. 4 we present a  simple example  of duality between 
description of motion of an ideal fluid in external gravitational field
and of dynamics of fluid with small viscosity in flat space.
In Sect. 5 we argue that the chiral  
anomaly results in unique predictions
for $D_{\alpha}j^{\alpha}_{5}$, where $j^{\alpha}_5$ is the total axial current,
see, e.g., Eq. (\ref{chve}), and $D_{\alpha}$ is the covariant derivative
in external gravitational field. Sect. 6 is devoted to a simplified treatment
of  fluid instabilities.

\section{Conservation laws in non-dissipative limit}

In this section we introduce various
axial currents, or helical motions and 
review conservation laws of chiral hydrodynamics in
the non-dissipative limit.  

Let us start with a well-known remark that the chiral 
anomaly can be cast into a form of
conservation of a generalised chiral charge which 
is non-vanishing for helical  
configurations of magnetic field:
\begin{eqnarray}\label{nonnaive}
Q_5^{conserved}~=~Q_5^{naive}~+~\frac{e^2}{4\pi^2} \mathcal{H}_{magn~hel.}~~,\\
\mathcal{H}_{magn~hel.}~=~\int d^3x\vec{A}\cdot \vec{B}~~,\nonumber
\end{eqnarray}
where
$\vec{A}$ is the vector potential of electromagnetic field and $\vec{B}$ 
is the magnetic field
and $\mathcal{H}_{mgn~hel}$ is the so called magnetic helicity.

At this point, there is no obvious place yet for the chiral vortical current,
see Eq (\ref{chve}). But 
there do exist various ways to appreciate the emergence
of a unified chiral current, like (\ref{chve}).
In particular, generalisation of (\ref{nonnaive}) to  the hydrodynamic case 
can be obtained \cite{shevchenko} through the 
substitution:
\begin{equation}\label{substitution}
eA_{\mu}~\to~eA_{\mu}+\mu u_{\mu}~,
\end{equation}
where $\mu$ is the chemical potential associated with the charge $e$. 
 Eq. (\ref{substitution}), in turn, 
is substantiated on
the grounds of similarity between the chemical-potential term
in the Hamiltonian, 
$\delta H~=~-\mu\cdot Q$
and the 
electromagnetic interaction with external field $\delta H~=~-ej^{\mu}_{el}A_{\mu}$. 
If we treat both terms perturbatively  then Eq. (\ref{substitution})
seems quite obvious.

Upon using (\ref{substitution}), one arrives at the expression of a 
quantum-mechanically conserved
axial charge suited for applications to hydrodynamics:
\begin{eqnarray}\label{totalcharge}
Q_5^{hydro}~=~Q^5_{naive}~+~\frac{e^2}{4\pi^2}\mathcal{H}_{magn~hel.}~+~\\
\frac{1}{4\pi^2}\mathcal{H}_{fluid~hel}~+~\frac{e}{2\pi^2}\mathcal{H}_{fluid-magn~hel}~~,\nonumber
\end{eqnarray} 
where the fluid helicity $\mathcal{H}_{fluid~hel}$ and fluid-magnetic 
helicity $\mathcal{H}_{fluid-magn~hel}$
are defined as
\begin{equation}\label{fluidhelicity}
\mathcal{H}_{fuid~hel}~=~
\int d^3x \epsilon^{0ijk}(\mu u_i)\partial_{j}(\mu u_k)~~,
\end{equation}
and
\begin{eqnarray}\label{fluid-magnetic}
 \mathcal{H}_{fluid-magn~hel}~=~\int d^3x\epsilon^{0ijk}(\mu u_i)\partial_jA_k 
~=\\ \nonumber
\int d^3x\epsilon^{0ijk}(A_i)\partial_j(\mu u_k) .
\end{eqnarray}
Note that all the kinds of helicities entering (\ref{totalcharge}) were introduced first
long time ago in the context  of magneto-hydrodynamics,
 see, e.g., \cite{bekenstein} and references therein.

One might question the validity of (\ref{substitution}) since
it is well known that
introduction of the chemical potential does not result in any extension
of the chiral anomaly. 
The famous triangle graph, with external electromagnetic legs is known
to be the only source of non-conservation of the original axial charge
$Q_{naive}$. 
And Eq. (\ref{substitution}) is, at first sight, in contradiction
with this assertion. 

To address this question, turn
to the specific case of ideal fluid. Then, the central point is that
various terms in the r.h.s. of Eq. (\ref{totalcharge}) can be separately
conserved on equations of motion for ideal fluid.
This is simplest to demonstrate in case of the magnetic helicity
entering Eq. (\ref{nonnaive}).
Indeed, kinematically we get:
\begin{equation}
\frac{d}{dt}\mathcal{H}_{magn~hel}~=~\int d^3x (\vec{E}\cdot \vec{B})~,
\end{equation}
where $\vec{E}$ is the electric field. 

However,
as is commonly known,  electric field is completely screened
inside an ideal conductor, $\vec{E}~=~0$.
Indeed, in general case of finite conductivity $\sigma_E$ electric current is
given by $\vec{j}_{el}~=~\sigma_E\vec{E}$. Since the current is to stay finite also in
the limit of the conductivity tending to infinity, $\sigma_E~\to~\infty$, one
concludes that 
\begin{equation}\label{ideal}
\lim_{\sigma_E~\to~\infty}{\vec{E}}~\to~0
\end{equation}
 in this limit. 
Thus, we come to a paradoxical conclusion that the chiral anomaly
resulting in $\partial_{\mu}j^{\mu}_5~\sim~e^2(\vec{E}\cdot \vec{B})$
does not signal any non-conservation of axial charge
if one evaluates matrix element of it 
over a state of ideal fluid in equilibrium.

Clearly enough, the conservation of the magnetic helicity in case of ideal liquid,
just discussed, is of a different nature than, say, conservation of the original
axial charge $Q_{naive}$ in field theory. 
The magnetic helicity does not correspond, 
at least within the derivation given above,
to any Noether charge. Its conservation is not a manifestation of any symmetry
of the underlying Lagrangian but is rather of dynamical nature.
 
 The condition of vanishing of the electric field 
inside an ideal conductor is modified
if one allows for variations of chemical potential $\mu$ and/or temperature $T$.
In this, more general case one has
$$\vec{j}_{el}~=~e\sigma_E\Big(\vec{E}-T\vec{\nabla}(\mu/T)\Big)~.$$
The condition (\ref{ideal}) for the fluid being ideal is 
changed respectively: it is the combination of 
the two terms in the right-hand side which is to vanish. 

This change is immediately reflected in the helicities conservation laws.
Namely, for  the divergence
of the fluid-helicity current (\ref{fluid-hel}) one gets:
\begin{equation}\label{derivative}
\partial_{\alpha}j^{\alpha}_{fluid~hel}~=~\frac{T^2\mu s}{(p+\epsilon)}\omega^{\alpha}
\partial_{\alpha}\Big(\frac{\mu}{T}\Big)~,
\end{equation} 
where $p$ and $\epsilon$ are pressure and energy density, respectively, $s$ is entropy density.
It is clear now
 that in case $\mu\neq 0$ (so that the fluid is charged) and
absence of external fields ($\vec{E}=0$) the fluid helicity is conserved,
$\partial_{\alpha}j^{\alpha}_{fluid~hel}=0$, in the dissipation-free limit.
This is, probably, the simplest case when imposing the dissipation-free limit 
brings in a new conservation law. And in the next section 
we will mostly concentrate on this
case. 
If both external electromagnetic fields and gradient of the $\mu/T$ ratio
are present, it is the combination of the three
helicities entering the r.h.s. of Eq. (\ref{totalcharge}) which
is conserved for ideal fluid 
\footnote{If one includes temperature-dependent terms as well, 
there is one more current, proportional  to $T^2$
which is conserved in the dissipation-free limit, see, e.g., \cite{golkar,avdoshkin}.}.

Within the approach outlined above, the fundamental field theory is invoked only 
on the level of derivation of the chiral anomaly. 
There exists a more ambitious field theoretic approach to hydrodynamics
which introduces universal infrared degrees of freedom
relevant to any fluid, see, e.g., recent papers
 \cite{nicolis} and references therein. Essentially, these 
infrared degrees of freedom are
deviations of the coordinates of an element of fluid from
their equilibrium values. Typically, one introduces scalar fields $\varphi^I$
with non-trivial expectation values in the equilibrium:
\begin{equation}
<\varphi^I>~=~x^I ~,
\end{equation}
where $x^I$ are the equilibrium positions of coordinates of an element of fluid.
There are certain symmetries imposed on the interaction of the fields $\varphi^I$
and invariants can be ordered according to the number of derivatives from
the fields. The lowest-order invariant is $B~\sim~\Big(\epsilon^{\alpha\beta\gamma\delta}
\partial_{\alpha}\varphi^I\partial_{\beta}\varphi^J\partial_{\gamma}\varphi^K\epsilon_{IJK}\Big)^2 
 $. Respectively, the field theoretic action density in this approximation is given 
in terms of a function $f(B)$ of the invariant $B$:
\begin{equation}\label{invariance}
S_{hydro}~=~\int d^4x f(B)~~.
\end{equation}
Starting from this action one can develop the Hamiltonian formalism and so on.

Yet another approach utilises variational approach and Hamiltonian formalism,
for a recent presentation and further references see \cite{abanov}. 
Again, all the effective infrared degrees of freedom of an ideal liquid
are included into consideration.  One demonstrates \cite{abanov}  that
the ``naive'' axial charge and the fluid helicity are separately conserved,
modulo the chiral anomaly.

To summarise, in case of ideal fluid there are extra conservation laws 
protecting various helicities. Account for the anomaly in presence of
external electromagnetic fields unifies all kinds of 
classically conserved axial charges into a single
charge (\ref{totalcharge}) conserved on the quantum level. This beautiful picture
is challenged, however, upon inclusion of non-vanishing
 viscosity.

\section{Motivation to introduce effective gravitational field}

It is known since long, see, e.g., \cite{bekenstein}, that, say,
 fluid helicity (\ref{fluidhelicity})
is no longer conserved once non-vanishing viscosity is accounted for. 
Indeed, in most cases dissipation brings fluid into  the equilibrium state which 
is  nothing else but the whole of the fluid at rest in a particular frame. 
In this, trivial state the fluid helicity disappears
independent of its initial value
\footnote{It is actually less trivial that there exist
motions corresponding to non-vanishing chiral effects
and consistent with viscosity $\eta\neq 0$, see \cite{bhattacharaya}.}.
The conclusion that conservation of the hydrodynamic axial charge
(\ref{totalcharge}) is inconsistent with $\eta\neq~0$ sounds puzzling,
since it implies, at first sight, that viscosity cannot be triggered by a
chiral invariant interaction \cite{avdoshkin}. In these notes, we 
are looking primarily for a resolution of this paradox.
  
To get a better insight into the problem, let us start with approximation
of ideal fluid and then ``switch on'' a small shear viscosity $\eta$. Furthermore,
assume that there is a region
where the fluid is rotating with angular velocity ${\vec{\Omega}}$. 
Then the contribution of this region into the fluid helicity is given by:
\begin{equation}\label{example}
\mathcal{H}_{fluid~hel}~\sim~\mu^2\int d^3x~ {\vec{ v}(x)}\cdot{\vec {\Omega}}~~,
\end{equation} 
 and we assume that it is non-vanishing. In equilibrium and in 
absence of the viscosity we  have 
$\vec{v}(x)$ constant as a function of time.
Thus, the charge (\ref{example})
is conserved as a consequence of the equation of motion of ideal liquid
which, under the simplifications specified, reads as
$$\frac{d\vec{v}_{\parallel}}{dt}~=~0~,$$
where $\vec{v}_{\parallel}$ is the velocity of the fluid parallel to the
vector of angular velocity $\vec{\Omega}$. Note that this equation of motion is not specific
for chiral fluids.

Include now a (small) shear viscosity. Rotation is consistent
with non-vanishing viscosity and, therefore, for variation with time of the
fluid helicity we get:
\begin{equation}\label{estimate}
\frac{d }{dt}\mathcal{H}_{fluid~hel}~
\sim~\mu^2\int d^3x~ {\vec{ a}_{fluid}}\cdot {\vec {\Omega}}~,
\end{equation} 
where $\vec{a}_{fluid}$ is the acceleration of the element of fluid considered. 
 In general, the fluid helicity is no longer conserved.
 Indeed, one can have the product
${\vec{ v}}\cdot {\vec {\Omega}}$ non-zero but 
dependent on the distance $\rho$ from the axis of the rotation. 
Then the viscosity $\eta\neq~0$
induces non-conservation of $\mathcal{H}_{fluid~hel}$, see
(\ref{estimate}).

Our central point is that
the estimate (\ref{estimate}) can be viewed  as a hint 
that the non-conservation of the axial current (in presence of shear viscosity)
can be imitated by equations of motion of 
ideal fluid in  external gravitational field. Indeed,
according to the equivalence principle,
 the effect 
of going into an accelerated frame can be imitated by introduction of an
external gravitational field.
Therefore, in presence of gravitational field (and in the non-relativistic limit)
we expect that the equation of motion of the ideal liquid
would look as $$\frac{d\vec{v}}{dt}~=~\vec{a}_{grav}$$
where $\vec{a}_{grav}$ is the acceleration imposed by 
the external gravitational field. Choosing the gravitational field such that
$$\vec{a}_{grav}~=~\vec{a}_{fluid}$$
would reproduce (\ref{estimate})
and we could trade viscosity for a certain external gravitational field.

Note that  in our case, the acceleration $\vec{a}_{fluid}$
entering Eq. (\ref{estimate}) 
is determined by dynamics of (viscous) fluid.
Thus, we  introduce an ``effective'' gravitational field,
 to reproduce the same acceleration
$\vec{a}_{fluid}$. Mathematically, our hypothesis can be formulated as Eq. (\ref{crucial})
and we will discuss it in length in the next section. 

Let us mention that it was argued first 
long time ago by Luttinger \cite{luttinger} that introduction of a gravitational
potential $\varphi_{gr}$ can be useful for hydrodynamic studies.
Namely, to imitate the effect of the temperature gradient
one can introduce an effective gravitational potential $\Psi$
coupled to energy density $\mathcal{E}$ through the Hamiltonian:
\begin{equation}
H_L~=~\int d^3x \Psi \mathcal{E}~~,
\end{equation}
where the potential $\Psi$ is adjusted to balance the thermal force,
$\nabla \Psi~=~\nabla T/T$.
  For recent applications of this idea to
dynamics of chiral fluids see \cite{stone,zahed}.  Reduction of the problem
with a non-vanishing gradient of temperature to
 the problem of motion in an external gravitational
field allows to use mucroscopic language to describe
thermal effects.
Similarly, we are proposing to mimic the effect of (small) shear viscosity
by a gravitational field. This analogy allows us to use well known 
equations from the general relativity.

Remarkably enough, it was argued very recently, see \cite{liu} and references 
therein, that consideration of dissipative fluids within field-theoretic approach 
to the general theory of fluids 
brings in the notion of ``emergent spacetime''. 
It seems plausible that the metric tensor introduced in \cite{liu}
to describe the emergent spacetime 
coincides with the ``effective''  gravitational field 
introduced above. However, such an identification is not crucial for our purposes here and 
detailed discussion of the issue is beyond the scope of the present notes.

\section{Classical gravitational ``anomalies''}

 Thus, we turn to considering motion in an external gravitational field
which induces acceleration on matter. Moreover, for our purposes now
there is no need to distinguish between  ``genuine'' and ``effective''
gravitational fields and we will drop the index ``effective'' from the metric tensor
for the moment.


Imagine that in the flat-space limit there exists a conserved current,
$\partial_{\mu}j^{\mu}~=~0$. Then, in presence of external gravitational field
the ordinary derivatives $\partial_{\mu}$ are replaced by  covariant ones, $D_{\mu}$:
\begin{equation}
\partial_{\mu}j^{\mu}~=~0~\to~D_{\mu}\mathcal{J}^{\mu}~=~0~,
\end{equation}
where $\mathcal{J}^{\mu}$ transforms as a vector under general coordinate transformations.
Note that in the limit of vanishing gravitational 
field the current $\mathcal{J}^{\mu}$ coincides with
the current $j^{\mu}$ of special relativity. However, $\mathcal{J}^{\mu}$
might contain terms of first order in gravitational potentials (which
are uniquely fixed by the requirement that $\mathcal{J}^{\mu}$
is transformed like a vector under general coordinate transformations).

For an arbitrary vector $\mathcal{J}^{\mu}$ and in standard notations, 
see \cite{landau}, one has: 
\begin{equation}
\mathcal{J}^{\mu}_{;\mu}~=~\frac{\partial \mathcal{J}^{\mu}}{\partial x^{\mu}}+
\Gamma^{\mu}_{\mu\rho}\mathcal{J}^{\rho}~,
\end{equation}
where $\mathcal{J}^{\mu}_{;\mu}\equiv D_{\mu}\mathcal{J}^{\mu}$ and 
$\Gamma^{\mu}_{\nu\rho}$ are Christoffel symbols. Moreover,
\begin{equation}\label{meaning}
(\mathcal{J}^{\mu})_{;\mu}~=
~\frac{1}{\sqrt{-g}}\frac{\partial(\sqrt{-g} \mathcal{J}^{\mu})}{\partial x^{\mu}}
\end{equation}
where $g$ is the determinant constructed on the metric tensor $g_{\mu\nu}$.

The physical meaning of (\ref{meaning}) is of course absolutely transparent. 
Indeed,
in  presence of a gravitational field
an element of physical volume is written as $dV^{(4)}_{phys}~=~\sqrt{-g}d^4x$
and Eq (\ref{meaning}) corresponds to conservation of the number of particles
in a unit physical volume, $dn~=~n(x)dV^{(3)}_{phys}$, where $n(x)$ is the density
of charged particles, the same as in, say, Eq. (\ref{chme}).
However, if one  treats the gravitational field as
a usual field in the flat space-time ${x^{\mu}}$, then Eq. (\ref{meaning}) looks
so as if the current $j^{\mu}$ is no longer conserved:
\begin{equation}\label{exact}
\partial_{\mu}\mathcal{J}^{\mu}~+~(1/2)\mathcal{J}^{\mu}\partial_{\mu} (\ln (-g))~
=~0~~.
\end{equation}
Note that Eq.(\ref{exact}) is exact classically.  

If there is a gravitational chiral anomaly then one has
\cite{alvarezgaume}:
\begin{equation}\label{anomaly}
D _{\mu}\mathcal{J}_5^{
\mu}~= ~C_{gr}
R_{\alpha\beta\gamma\delta}\tilde{R}^{\alpha\beta\gamma\delta}~~,
\end{equation}
where $R_{\alpha\beta\gamma\delta}$ is the Riemann tensor
and $C_{gr}$ is a constant. 
The r.h.s. of (\ref{anomaly})  
of   second order in gravitational field while the effect of acceleration,
discussed above, 
can be imitated by an external field treated as a first-order correction.
Thus, for our purposes we can neglect the r.h.s. of (\ref{anomaly}).

Note that the chiral anomaly
is quite commonly quoted as containing an ordinary derivative 
in the l.h.s., $\partial_{\mu}j^{\mu}_5~=~C_{gr}R\tilde{R}$.
Such a form assumes perturbative evaluation of the matrix element 
of the axial current with emission/absorption of gravitons
in flat space-time. Eq. (\ref{anomaly}),
on the other hand, refers to a general, $x$-dependent gravitational field,
or non-vanishing condensate of the graviton field,
\begin{equation}\label{hmunu}
<h_{\mu\nu}>~\neq~0
\end{equation}
 where
$h_{\mu\nu} \equiv~g_{\mu\nu}(x)-\eta_{\mu\nu}$.
We will come back to continue this discussion later.

Let us now apply (\ref{exact}) to the case of weak gravitational field, 
$|<h_{\mu\nu}(x)>|~\ll~1$.
Then we get:
\begin{equation}\label{weak}
\partial_{\mu}\mathcal{J}^{\mu}~\approx~-(1/2)\mathcal{J}^{\mu}\partial_{\mu}h~~,
\end{equation}
 where $h~\equiv~h_{00}-\Sigma_ih_{ii}$.
Moreover, the $h_{00}$ component is fixed by the equivalence principle:
\begin{equation}\label{h00}
g_{00}-1~\approx~2\phi_{gr}~\approx~-2\vec{a}\cdot \vec{x}~.
\end{equation}
where $\phi_{gr}$ is the Newtonian potential and $\vec{a}$ is the acceleration.

As is well known, the components $h_{ii}$ are not fixed by the equivalence principle
and  gauge dependent. In other words, 
the result of calculation is  locally sensitive to the gauge fixing,
if the problem is not entirely governed by non-relativistic physics.
The best known example of this type is deflection of light by the sun. Keeping 
only $h_{00}~\neq~0$ underestimates the effect by a factor of 2 in this case.
Using the gauge
\begin{equation}\label{gauge}
h_{11}~=~h_{22}~=~h_{33}~=~h_{00}
\end{equation}
 restores the lacking factor of 2 locally.  (A small deflection
angle is evaluated by integrating the effect of gravity over the unperturbed
light trajectory in flat space. Only the integral 
over the whole trajectory is gauge independent. ``Locally'' means that
the correct numerical factor is reproduced already prior to the
integration). Thus, we will explore also the gauge (\ref{gauge}).
 
With Eq. (\ref{exact}) in hand, we can come back  to 
the example of motion described by Eqs. (\ref{example}), (\ref{estimate})
and check  whether it actually fits Eq. (\ref{exact}).
For this purpose, we compare results of two calculations.
First, we evaluate divergence of the fluid-helicity current for 
the fluid configuration (\ref{example}) and then compare the result with
Eq. (\ref{weak}).

Let us fix first notations. 
We consider the following (non-relativistic)
 fluid configuration:
\begin{eqnarray}
u^{(1)}~=~(1/2)\Omega \cdot x^{(2)}~,~~u^{(2)}~=~-(1/2)\Omega\cdot x^{(1)}~,\\ 
\nonumber
~u^{(3)}~\equiv~u^{(3)}(t)~,
\end{eqnarray}
where $\Omega$ is the angular velocity considered to be
a constant,$u^0, u^{(i)} ~(i=1,2,3)$ are components of
the 4-velocity, $u^{(0)}~\approx~1+1/2(u^{(1)})^2+1/2(u^{(2)})^2+1/2(u^{(3)})^2$. 
Furthermore, 
we are evaluating the current 
$j^{\mu}_{fluid~hel}$ introduced in eq. (\ref{fluid-hel}). The notations are such than
that 
$\epsilon^{0123}~=~1$ and the minkowskian metric tensor is chosen to be 
$g_{00}=1, g_{ii}=-1$. Also, for the purpose of the present exercise the
chemical potential is kept constant. 

 
For the components of the current $j^{(i)}\equiv~\mu^{-2}j^{(i)}_{fluid~hel}$ we find:
\begin{eqnarray}
j^{(1)}=-(1/2)\Omega x^{(1)}a^{(3)}~, j^{(2)}=-(1/2)\Omega x^{(2)}a^{(3)},\\ \nonumber
~j^{(3)}=-\Omega,
\end{eqnarray}
where $a^{(3))}\equiv d u^{(3)}/dt$. And, finally, for the divergence of the current
we get:
\begin{equation}\label{expected}
\mu^{-2}\partial_{\mu}j^{\mu}_{fluid~hel}~=~-2a^{(3)}\Omega~~,
\end{equation}
where the acceleration $a^{(3)}$ is determined by dynamics of
the viscous fluid. We expect that this relation can be reproduced  
by considering ideal fluid in the effective gravitational field.

The curved-space counterpart of the fluid-helicity current is given by:
\begin{equation}
\mu^{-2}(j^{\mu}_{fluid~hel})_{cov}~=
~\big(-g\big)^{-1/2}\epsilon^{\mu\nu\alpha\beta}u_{\nu}D_{\alpha}u_{\beta}~.
\end{equation}
The condition of vanishing of the covariant derivative (\ref{meaning}) becomes
\begin{equation}\label{result}
\partial_{\mu}\Big(\epsilon^{\mu\nu\alpha\beta}u_{\nu}D_{\alpha}u_{\beta}\Big)~=~0~.
\end{equation}
Using Eqs (\ref{h00}), (\ref{gauge}) with $\vec{a}~=~(0,0,a^{(3)})$
we arrive to the same Eq. (\ref{expected}), as expected.  

Two
 comments concerning this result are now in order.
First, Eq. (\ref{result}) is sensitive only to the component 
$h_{00}$ of the weak gravitational field. This is not obvious apriori since
the divergence of the fluid-helicity current is equally contributed by
the zeroth and spatial components of the current, while non-relativistically
we could expect dominance of the zeroth component, see discussion in Sect. 2.
Our second comment is that effect of an external gravitational potential
on the chiral currents was evaluated in Ref. \cite{zahed}. In this paper,
the authors consider the Fermi-sphere model for the chiral fluid and
evaluate the axial current starting from first principles. 
In our approach, the chiral vortical current (\ref{chve})
arises quantum mechanically while the last step, that is inclusion 
of the gravitational potential, is an application of a classical
equation (\ref{meaning}). Algebraically, results appear to coincide 
while their interpretation is different \cite{minkin}.

 It is worth empasising  that  presence of linear 
in the acceleration $\vec{a}$ terms
is
specific for chiral effects. 
In case of 
the most standard hydrodynamic current,
$j^{\mu}~=~nu^{\mu}$, the 4-vectors of the velocity and acceleration
are orthogonal to each other, $u^{\mu}\partial_tu_{\mu}=0$ because of
the normalisation condition, $u^{\mu}u_{\mu}=~=~-1$.
As a result the r.h.s. of Eq. (\ref{weak}) vanishes in this case
(at least, for a constant $n$) .

To summarise this section,
 for a particular example  considered we can reinterpret 
the non-conservation of  the fluid helicity due to dissipation
as a change of geometry of space
due to presence of an external gravitational field.
The gravitational field itself is an implicit
 function of initial distribution
of velocities and masses and of viscosity of the fluid.
The basic limitation of the example considered
is that it is linear in the effective gravitational field
 triggered by viscosity. 
Such an approximation can be justified by smallness
of viscosity. On the other hand, on physical grounds one can expect
that the relation between dissipation and geometry, encoded in Eq. (\ref{crucial})
remains true also in higher orders in the gravitational field.

\section{Remarks on higher orders in gravitational field}


In this section, we will clarify a few points concerning covariant 
conservation of the axial current.
Our remarks are algebraic in nature
and apply both in case of ``genuine'' and ``effective''
gravitational fields, and we do not distinguish between these two cases
in this section.

Let us go back to evaluation of divergence of axial current in
gravitational background. For simplicity, we will not introduce
external electromagnetic fields. In this approximation
and  in flat space the axial current
is given by (\ref{chve}). Next, we inroduce a non-trivial 
position-dependent background
 $<h_{\mu\nu}>\neq~0$ where  $h_{\mu\nu}\equiv g_{\mu\nu}-\eta_{\mu\nu}$
and evaluate divergence of axial current. Our results
of the preceding section  can be summarized 
in the following way:
\begin{equation}\label{anomalyrelated}
\Big(\partial_{\alpha}~j^{\alpha}_5\Big)_{<h_{\mu\nu}>\neq 0}~=~
C\partial_{\alpha}\epsilon^{\alpha\beta\gamma\delta}
(\mu u_{\beta})D_{\gamma}(\mu u_{\delta})~+~C_{gr}R\tilde{R}
\end{equation}
where $C_{gr}R\tilde{R}$ is the standard gravitational chiral anomaly.
The central point is that the r.h.s. of  Eq (\ref{anomalyrelated})
starts with a term which is proportional to the coefficient $C$
which  is a label for the (``electromagnetic'') chiral anomaly.

Let us remind the reader, how it happens, on the technical side,
 that divergence of a matrix element of the axial
current over a pure gravitational background becomes sensitive to
the chiral anomaly associated with electromagnetic interactions. 
Vanishing of the covariant derivative of a current, $D_{\alpha}j^{\alpha}_5=0$ 
is a textbook 
formulation of 
generalization of a 
(non-anomalous) conservation law to the case of gravitational background.
Generally speaking, the divergence of the axial current $\partial_{\alpha}j^{\alpha}_5$
could receive at this step various contributions.
However, we found out that, in our simplified case there is in fact
a single term
proportional to the anomaly-related coefficient $C$.

To elucidate the origin of this factor of $C$
 concentrate on  the substitution (\ref{substitution}).
To justify it one starts with the perturbative Lagrangian
$$\delta L~=~eA_{\mu}j^{\mu}_{el}~+~\mu u_{\mu}j^{\mu}_{el}~,$$
where the second term in
the r.h.s. is the hydrodynamic generalisation of $\delta L~=~- \delta H~=-\mu\cdot Q$.
Using  the language of Feynman graphs
we come to a hydrodynamic analogy of the chiral anomaly
where $eA_{\mu}$ is replaced by $\mu u_{\mu}$.
This is the origin of the fluid-helicity current $j^{\alpha}_{fluid~hel}$
in Eq. (\ref{chve}) and this is how the coefficient $C$ enters
the expression (\ref{chve}).

The fluid-helicity current is not kinematically conserved, 
$\partial_{\alpha}j^{\alpha}_{fluid~hel}\neq 0$. This 
property is inherited from the fundamental chiral anomaly
since the $j_{fluid~helicity}$ term in
(\ref{chve}) is generated by the triangle graph, with two $\mu u_{\alpha}j^{\alpha}_{el}$
vertices.
However, the fluid helicity is conserved on the equations of motion of 
ideal fluid (see Sect. 2 for discussion). And this is a manifestation
of the well-known theorem that introduction of chemical potentials (and temperature)
does not affect the fundamental triangle anomaly,
in the sense that there are no further matrix elements
which would exhibit  non-conservation of chirality. Moreover, 
the ideal-fluid approximation
is essential since only then theory of  fluids
 can consistently be reduced to a unitary field theory, see discussion in Sect. 2.
That is why analysis of the ideal fluid in the language of Feynman graphs 
is granted to obey general theorems of field theory.
Thus, Eq. (\ref{chve})  is consistent both with substitution (\ref{substitution}) and 
the statement that the non-conservation of
the total axial charge (\ref{totalcharge})
is entirely determined by the 
original triangle graph.  

Next, switch on gravitational field (still within the ideal-fluid approximation).
Then for the axial current of the fundamental theory 
and to first order in gravitational interaction we have $\partial_{\alpha}j^{\alpha}_5=0$
and can use here the ordinary derivative instead of the covariant derivative.
This is a well-known
manifestation of chirality conservation by gravitational interaction
of (fundamental) massless fermions. However, if we generalise conservation
of the fluid-helicity current
to the presence of external gravitational fields
 we have to use the general relation (\ref{crucial}) 
and keep the covariant derivative, not 
replacing it by ordinary derivative. This is because the
chiral vortical current in Eq. (\ref{chve}) represents
a hydrodynamic analogy of the fundamental triangle graph.
In this way we can trace the origin of the anomaly-related coefficient $C$
in the r.h.s. of Eq. (\ref{anomalyrelated}).

  In Sect. 2 we mentioned similarity between calculations in external gravitational field
and coordinate-dependent condensate of a scalar (Higgs) field. In conclusion
of this section we elaborate further on this remark.
Let us remind the reader that currents induced by coordinate-dependent 
condensates were introduced by Goldstone and Wilczek \cite{goldstone}. 
As is observed in \cite{harvey}
the electromagnetic current
\begin{equation}\label{harvey}
j^{\rho}_{el}~=~(Const)\epsilon^{\rho\mu\nu\sigma}\partial_{\mu}F_{\nu\sigma}
\end{equation}
is in fact not conserved if there is a vortex (such that the vacuum
 expectation value of $\phi$ is not zero in the vacuum, $<\phi>\neq 0$,
but vanishes along a line $<\phi (x)>_{x=0}=0$). In this case
\begin{equation}\label{divergence}
\partial_{\rho}j^{\rho}_{el}~\sim~\delta^{(2)}(x_{\perp}=0)~~.
\end{equation}
Note that the current (\ref{harvey}) is a non-anomalous one-loop effect.

The singular nature of the divergence of the current (\ref{divergence})
implies  that the current (\ref{harvey}) cannot be
 treated classically and there is production of particles. 
The particles correspond to zero modes of massless fermions
 in the background of the vortex field,
and production of particles by electric fields 
corresponds to the chiral anomaly in (1+1) dimensions. 

A natural question arises whether there is a similar effect
in the gravitational case, when $<h_{\mu\nu}>\neq 0$.
Eq. (\ref{meaning}) might indicate that the classical treatment fails
if the determinant $g~\to 0$. The determinant
$g$ vanishes near the horizon for a Rindler observer. 
Indeed, the metric induced  is given by $ds^2~=~r_c^2dt^2-(dx^i)^2$,
with $r_c\to 0$ at the horizon. The physics in Rindler space, in turn, 
is similar to physics
of a large black holes near the horizon. Thus, basing on 
Eq. (\ref{meaning}) one could speculate that
production of particles and quantum anomaly are  relevant in case $g~\to~0$
\footnote{Note that while the determinant $g$ is gauge dependent 
its vanishing is gauge 
independent.}.
And there is, indeed, a deep connection between the Hawking radiation 
from the horizon 
and chiral anomaly in (1+1) dimensions, as is
discovered by authors of papers in Ref. \cite{wilczek}.

\section{Phase transitions}
\subsection{Instabilities of ideal fluids}

Phase transitions reveal structure of conservation laws
and of effective degrees of freedom
in a dramatic way. 
In this section we present an oversimplified version of ``theory''
of instabilities, or phase transitions of chiral fluids. 
One of such instabilities is a spontaneous production of helical configurations
of  magnetic field from chirally asymmetric plasma, $\mu_5\neq 0$.
In other words, chirality of microscopic degrees of freedom, elementary
fermions, is transformed into helicity of macroscopic magnetic fields.
Theory of this instability has been elaborated in quite great
detail, see, in particular, \cite{redlich,khaidukov} and we do not have anything to add
to this theory. 

We are rather interested in 
theory of another type of phase transitions
which is the transfer of chirality of elementary constituents to
macroscopic, helical motion of the fluid. Discussion of this phase
transition and of its possible astrophysical applications
is very recent \cite{avdoshkin,yamamoto}. 
First-principles theory
of this transition is 
much less developed.  

The reason is that  instabilities with
respect to generation of magnetic fields can be studied within 
(quantum) electrodynamics while field theory for ideal fluids
is developed only in case of small fluctuations around the equilibrium,
see the references in the Introduction. Therefore, spontaneous production
of vortices can hardly be considered analytically. 
Studies of vortices are mostly numerical.
A very interesting numerical result which might be relevant to our discussion
is obtained in Ref. \cite{burch}. One starts with effective action for 
ideal fluid (\ref{invariance})  which is treated thermodynamically, i.e. in
the Euclidean space. The action describes  thermodynamics of microscopic
degrees of freedom, either at finite temperature $T$ and entropy density $s$
or at finite chemical potential $\mu$ and density $n$ \footnote{In more detail,
one considers ideal fluid in absence of vortices. The two descriptions are
equivalent or dual to each other. In the expression of free energy
one replaces $Ts$ by $\mu\cdot n$, or vice verse \cite{nicolis}. }. 
Numerical simulations indicate phase transition to vortical states at
some temperature (or, respectively, chemical potential).

We will present simple estimates of characteristic energies and momenta
of effective degrees of freedom
associated with the instabilities \cite{kirilin1}. 
These estimates can be confronted with
much more quantitative calculations  \cite{redlich} referring
to spontaneous decay of chiral plasma into helical magnetic fields.
The estimates turn to be correct, within approximations made.
Then we apply similar estimates to the process of
decay into fluid vortices which is understood much worse analytically
(see remarks above). The estimates indicate
that straightforward application of hydrodynamic expansions (in derivatives)
underestimates infrared effects. Moreover, decay
to a vortical state looks plausible.
Finally, we suggest an interpretation of the phase transition
in terms of the effective gravitational field introduced in Sect. 3.

\subsection{General  mechanism of instability}

  A crucial point for a phase transition to take place is energy balance.
Conservation of the total axial charge provides an extra constraint on
the dynamics of the instabilities.

We have the following problem to consider.
There are, say, two 
 classically conserved charges,
$Q_1,Q_2$ which are mixed up into a single conserved charge 
$Q_{total}=Q_1+Q_2$ because of a generic ``anomaly".  
The chemical potential associated with
$Q_{total}$ is assumed to be non-vanishing
and only transitions consistent with conservation of $Q_{total}$ are allowed.
We start with a state where the whole of the charge is equal to 
$Q_1$, so that
\begin{eqnarray}\label{initial}
(Q_{total})_{initial} ~=~Q_1\\ \nonumber
E_{initial}~=~\mu\cdot Q_1
\end{eqnarray}
Imagine that degrees of freedom associated with $Q_2$ get excited,
$Q_2=~|\delta Q_1|$ where the change of $Q_1$ is small $|\delta Q_1|\ll Q_1$.
Then it is a reasonable guess that the change in energy is
given by: 
\begin{equation}\label{quadratic}
\delta E~=~-\mu\cdot|\delta Q_1|~+~(const)(\delta Q_1)^2 ~,
\end{equation}
where the term proportional to $(\delta Q_1)^2$ corresponds to kinetic
energy of the excited degrees of freedom.
If Eq. (\ref{quadratic}) indeed holds, then phase transition is favoured at least
for small $|\delta Q_1|$. Moreover, it is quite obvious that the final state
would have charges $Q_1,Q_2$ of the same
order, $Q_1 \sim~Q_2$, unless there is a hidden large parameter 
inherent to the problem
\cite{avdoshkin}.

Our central point is that possibility  (\ref{quadratic}) is realised
in case of decay of chiral fluid both into helical magnetic field and 
vortices.

\subsection{Spontaneous production of helical magnetic fields}

Let us now approach the problem of chiral fluid instability,
within framework just outlined, in case when axial charge
can be approximated by a sum of two terms, see eq. (\ref{nonnaive}).
Thus, we have
$$Q_1\equiv~Q_{naive}~,~~~ 
Q_2 \equiv e^2/(4\pi^2) \mathcal{H}_{magn~hel}~.$$
We start with an
initial state such that $Q_{naive}\neq 0$ and there is no magnetic field.
The phase transition in point is a spontaneous generation of
magnetic fields from chiral plasma, see, in particular, \cite{redlich,khaidukov}
and references therein.

Imagine that a helical magnetic field is generated with charge  $\delta Q_2$.
From definition of the magnetic helicity we have in the momentum space:
\begin{equation}
\delta Q_2~\sim ~\alpha_{al}  p_B \vec{A}^2~`
\end{equation}
where $\alpha_{el}=e^2/4\pi$, $ p_B$ is a characteristic momentum
of the magnetic field (simplest helical magnetic field is a combination
of three standing waves), and $\vec{A}$ is the vector potential.

Generation of the magnetic field would cost energy density of order
\begin{equation}\label{magnetic field}
\vec{B}^2~\sim ~ p_B^2\vec{A}^2
\end{equation}
The phase transition is energetically favoured provided that
$ p_B^2~\leq~\mu\alpha_{el} p_B~$, or
\begin{equation}\label{main}
 p_B~\sim~\alpha_{el}\mu~.
\end{equation}
Thus, we see that Eq (\ref{quadratic}) does hold, as far as orders 
of magnitude are concerned.

We come also to a new point. Namely,  
Eq. (\ref{main}) exhibits a generic feature of all the hydrodynamic
instabilities of chiral fluids. Instabilities  arise from a far infrared
region, with a large correlation length. In case (\ref{main})
the correlation length is parametrically enhanced as $\alpha_{el}^{-1}$.

There is another important point worth mentioning. The estimate 
(\ref{magnetic field}) does not tell us which quantity (if any) is
``of order unit'', magnetic field, $ \vec{B}^2$ or vector potential,
$\vec{A}^2$. The correct answer seems to be that it is $\vec{B}^2\sim\mu^2$
which is of order unit. 
The estimate for $\vec{A}^2$ then reads as:
\begin{equation}\label{tobefulfilled}
\vec{A} ~\sim~r_{typical}\times \vec{B}~, ~or~|\vec{A}|~\sim~\alpha^{-1}_{el}\mu~,
\end{equation}
where $\mu$ is the chemical potential. 
It is amusing that we count $|\vec{A}|$ as being ``large'' and extract 
observable consequences from that. Despite of the fact that $|\vec{A}|$
is apparently non-gauge invariant. Nevertheless, the estimates do make sense since
the charge $\mathcal{H}_{magn~hel}$ is gauge invariant.

Combining (\ref{main}) and
(\ref{tobefulfilled}) we find out that after the phase transition 
\begin{equation}
(\vec{B})^2~\sim~\mu^2,~~q_2~\sim~\alpha_{el}|\vec{A}||\vec{B}|~\sim 
\mu^3~,
\end{equation}
where $q_2$ is the density of the magnetic helicity. Note that
the small parameter $\alpha_{el}$ is canceled from the expressions for the
energy and charge density after the decay of the original chiral plasma.
Qualitatively, this picture was advocated in the preceding subsection.

\subsection{Decay of chiral fluid into vortices}

Proceed now to the case:
\begin{equation}\label{secondcase}
Q_1~\equiv~Q_{naive},~~Q_2~\equiv~\mathcal{H}_{fluid~hel},~~
Q_{total}^{initial}=Q_1~.
\end{equation}
The instability to  be discussed  
is the decay of the chiral fluid into vortices \cite{avdoshkin, yamamoto} . 

Our central point is that estimates of energies and charges
in this case are very similar   to the case of decay 
of the plasma into helical magnetic fields. This similarity 
is not obvious but comes up naturally in the far-infrared region.

At first sight, the case (\ref{secondcase})
 is very different from the preceding one.
Indeed, the density of the fluid helicity 
is proportional to $$\mathcal{H}_{fluid~hel}~\sim~\vec{v}\cdot curl\vec{v}$$
and is quadratic in velocity $\vec{v}$ in the non-relativistic limit. 
Since the energy is also quadratic in $\vec{v}$ the condition
(\ref{quadratic}) is 
apparently not satisfied. Moreover, the fluid helicity is 
apparently suppressed
in hydrodynamic approximation by an extra power in the gradient expansion
(since we have $curl ~\vec{v}$).

However, all these objections 
to the possibility of the phase transition are invalidated
by infrared divergences inherent to hydrodynamics. 
As a result,
the actually relevant estimate of the fluid helicity is 
provided by our toy example, see Eq. (\ref{estimate}),
with vorticity $\Omega$ being of zeroth order 
of smallness in the hydrodynamic approximation. 
Then, $\delta Q_1$ is of first order in the 
non-relativistic velocity $\vec{v}$ while the
energy is quadratic in $\vec{v}$. Thus, we have generically 
the same Eq. (\ref{quadratic}),
and the phase transition is favoured energetically.

The infrared divergences 
appear both within the framework
of perturbation theory  (for fluctuations near equilibrium)
and in terms of classical solutions for vortices.
As textbooks emphasise, 
vorticity classically costs energy tending to zero 
in the far infrared, or in the limit of large size 
of the vortex.
This follows from elementary estimates.
One can readily appreciate this point by 
 calling on  the analogy, suggested
by (\ref{substitution}), between  magnetic and vortical cases:
\begin{equation}
\big(\vec{A}~\sim ~\vec{B}\times \vec{r}\big)~\to
\big(\vec{v}_{\perp}~\sim~\vec{\Omega}\times \vec{r}\big)~~.
\end{equation}
We see that the component $v_{\perp}$ of the velocity
perpendicular to $\vec{\Omega}$ is ``infrared divergent'' 
because of an explicit
coordinate dependence.  
In other words, the vortex is described by a solution of
classical hydrodynamic equations.
Unfortunately, numerical estimates of the effect
of this infrared divergence are difficult to perform.
The reason is that 
vortical classical solutions are difficult to enumerate.
In other words, the phase space associated with the vortical solutions is poorly known.

In perturbation theory, to the contrary, the infrared divergences are
readily identifiable within the field-theoretic approach to ideal
fluids \cite{nicolis}  mentioned in the Introduction, see Eq. (\ref{invariance}).
On the other hand, perturbation theory does not tell us, what is the ultimate
configuration which the infrared instabilities drive the fluid to.

In more detail,
one expands in deviations $\pi^I(x)$ of positions of elements of fluid from
their  equilibrium values:
\begin{equation}
\phi^I~=~x^I~+~\pi^I~,
\end{equation}
where $I=1,2,3$, $\phi^I$ are scalar fields, $x^I$ are equilibrium positions,
or $x$-dependent vacuum expectation values in the language of 
the scalar fields, $<\partial_{\mu}\phi^I>~=~\delta^I_{\mu}$.
Moreover,
using action (\ref{invariance}) one can quantise the theory and
evaluate various correlator functions perturbatively. In particular, one finds \cite{nicolis}
for the Fourier transforms of
the propagators of $\pi^{'}$s in the limit of 
the frequency $\omega \to~0$:
\begin{eqnarray} \label{infraredd}
\lim_{\omega\to~0}{<\partial_i\pi^I,~\partial_j\pi^J>}~\sim~\\ \nonumber
\frac{P_T^{IJ}}{\omega_0}\frac{p^5}{\omega}~+~\frac{P_L^{IJ}}{\omega_0}p^3\omega
\end{eqnarray}
where $\omega_0$ is a constant, $P_{T}$ and $P_{L}$ are the transverse
and longitudinal projectors corresponding to the decomposition
$\pi^I~\equiv~(\partial^I\pi_L)/\sqrt{-\partial^2}+ \pi^I_T$.

Eq. (\ref{infraredd}) demonstrates clearly that at $\omega\sim p^3$
the correlation between variables $\pi_T^I$ becomes strong
and actually cannot be treated perturbatively. The origin of this infrared divergence is
the perturbative pole at $\omega = 0$. It is worth emphasising that the transverse
fluctuations $\pi^I_T$ correspond, in the language of perturbation theory,
to vortices. And the pole at $\omega =0$, see Eq. (\ref{infraredd}), is a manifestation 
of the same phenomenon of absence of barrier for creation of vortices
of large size, as discussed above.
 Infrared divergences of perturbation theory indicate
emergence of a classical solution, or a new vacuum state in the infrared 
region. Note that this ``classicalization'' seems to be
a general field-theoretic phenomenon, as argued recently  
in \cite{dvali}
(in connection with UV divergences).

Finally, appearance of the pole at $\omega=0$ (see Eq. (\ref{infraredd}))
can be traced back to the fact that quantisation 
is performed, as usual, in the quadratic approximation.
If one keeps non-linearities then one gets an estimate \cite{nicolis} 
\begin{equation}\label{observation}
\frac{\partial \Omega}{\partial t}~\sim~\eta p_{typ}^3\Omega~~,
\end{equation}
where $p_{typ}$ is a typical momentum
and $\eta$ is the viscosity.  Observation (\ref{observation})
can again be traced back to  the fact that creation of a vortex does 
not cost energy in the leading approximation.
Eq. (\ref{observation}) can also be used
to estimate corrections to our toy model
for leading fluid configurations in the infrared region, see Eq.  (\ref{example}).

To summarise, the cases of spontaneous production of 
macroscopic configurations of magnetic field and of 
vortices  have much in common {\it in the far-infrared region}. In both cases 
the instabilities are associated with far infrared.
Attempting to quantise excitations near the 
``naive'' equilibrium state (with $Q_1^{initial}=Q_{total}$)
brings to light strong interactions between excitations in some
regions of the phase space (see, e.g., \cite{khaidukov,nicolis}).
This inconsistency is apparently resolved by formation of 
coordinate-dependent
``condensate'' which is nothing else but classical solutions of the
corresponding differential equations. In case 
of the magnetic helicity one deals with
solutions of the Beltrami equation, while 
in case of helical motions one considers 
solutions of the Navier-Stokes equations. 
Processes in presence of coordinate-dependent
backgrounds are described by a kind of 
generalisation of the Callan-Harvey currents.  

\subsection{Emergent gravity?}

In the preceding subsection we argued that, dynamically,
decay of a chiral fluid into helical configurations of magnetic field 
and into vortices have much in common.
However, in case of magnetic fields we were able to derive also an
energy balance. A crucial point is that for a small variation $\delta Q_{naive}$
of the
charge the gain in energy is linear
in this small variation while the loss of energy is quadratic.
The loss of energy is associated with  
the energy  of the generated magnetic field. Note that as far as we consider external
magnetic fields, say, relation (\ref{nonnaive}), the energy of magnetic field
plays no dynamical role. But once we allow for instabilities,
 or generation of magnetic fields the energy 
 density $\epsilon~\sim~\vec{B}^2$ 
becomes crucial.
Moreover, we observed
that conservation of axial charge for viscous 
 fluids can be viewed as a modification of the naive charge
conservation due to presence of an effective gravitational field
$g_{\mu\nu}^{eff}$. Then the
non-conservation of the naive charge takes the form:
\begin{equation}\label{interference}
\partial_{\alpha}(j^{\alpha}_5)_{naive}~\sim~
\partial_0(q_5^{naive}) ~\sim~ (\partial_i g_{00}^{eff})\Omega~,
\end{equation}
where we kept the leading-order contribution, linear in the
gravitational  field, and $q^{naive}_5$ 
is the density of the naively conserved axial charge.

Eq. (\ref{interference}) looks as an analogy of, say,
fluid-magnetic  helicity (\ref{fluid-magnetic}).
Namely, we have a contribution to the axial charge
expressed as an integral over spatial coordinates.
The integrand is a product of a potential, or gauge-noninvariant term and of a
``gauge invariant'' term. The potential-type term is represented 
now by
a Christoffel symbol $\Gamma^i_{00}$, 
as it should be in case of gravitational field.
The vorticity $\vec{\Omega}$ is to be considered as a gauge-invariant
term, as is explained in the preceding subsection.

Thus, the expression for axial charge--with account of the emergent
gravitational field-- looks similar to the electromagnetic case. However,
the dynamics of phase transitions is governed also by 
the energy balance. In the examples we considered
the correction to axial charge was linear in a small parameter
while loss of energy is quadratic in the same parameter.
Thus, for the analogy to be held we are invited to speculate that
the effective gravitational energy
 contains also quadratic terms:
$$\epsilon_{gravity}~\sim~ \Gamma \Gamma~+...$$ The standard
expression for the energy 
of (fundamental)  gravitational fields
does have such terms. In our case, these quadratic terms would correspond
to the energy of ultra-violet degrees of freedom of the fluid.
The same is true in case of the electromagnetic 
decay of the chiral plasma. Namely, it is the
dynamics of the infrared degrees of freedom 
which drives the decay  while the dynamics at the ultra-violet scale
ensures stability at short times.

Thus, we hypothise that
the emergent gravitational field is to be treated as a dynamical one once
physics of 
phase transitions is included into the consideration. 
In other words, we are led to introduce
 emergent gravity.
Much more work is of course needed to make this hypothesis 
convincing.

  Note that
the phase transition we are discussing now (that is, decay of chiral plasma
into helical states of the effective gravitational field) would also
signify emergence of viscosity, even if one starts from an ideal fluid.
Phenomenologically, it would be of great interest  to check whether
the phase transition observed on the lattice
\cite{burch} results in emergence of a non-vanishing viscosity.

 \section{Conclusions}

We have reviewed conservation laws inherent to ideal fluids
emphasising the point that there are extra conserved currents
(apart from the Noether current of the underlying field theory).

We used this observation to suggest that introduction of viscosity
can be imitated by an emergent external gravitational field.
In more detail, we have considered the following construction.
We start with a state of ideal fluid in equilibrium and non-trivial
fluid helicity. Then we switch on shear viscosity, $\eta\neq 0$.
The corresponding dissipative force induces acceleration
which is a function of the initial distribution of velocities (and densities).
If we treat the problem in flat space, then the hydrodynamic charge
(\ref{totalcharge}) is no longer conserved. At least superficially,
this non-conservation is in variance with expectations
based on field theory.  
To elucidate the physical meaning
of this non-conservation we introduce gravitational potential which reproduces
the field of acceleration induced by viscosity.  
Then we demonstrate 
that 
the non-conservation of the charge under discussion
does correspond to
the classical equation (\ref{anomaly}) so that the would be non-conservation of charge
corresponds in fact to a change in the physical volume
induced by the effective gravitational field. 

From a more general perspective, we find out that field-theoretic formulation
of dissipative media assumes introduction of curved space. 
The metric is a function of the viscosity and of distribution of velocities.
So far, we could indicate the algorithm of evaluating the effective metrics
only to first order in viscosity. Note that our conclusion on
emergence of the curved space-time in description of dissipative media is in
accord with recent developments, see in particular \cite{liu}. 
The justification given above is, however,
is independent.

The approach discussed now is somewhat similar to the now-famous gauge-string
duality. In the latter case,  to describe dissipation one introduces an extra
(curved)
dimension. Propagation to the extra dimension corresponds to
a kind of ``disappearance'' from the physical space and describes dissipation. 
Within the approach considered here, curved space is also introduced
``everywhere". The metric tensor is determined by the fluid dynamics.
No further symmetry, like supersymmetry, is required from the
underlying fundamental theory at this stage.
In projection to the problem
of axial current, or helicity conservation,
the crucial effect is the change of the physical 4-volume as a function of the
emergent gravitational field. 
Note that in the field theoretic approach to   theory of
ideal fluid the freedom of the volume reparametrization
is a crucial element of the whole construction,
see Eq. (\ref{invariance}) and discussion around it.

According to the views presented here
the physical volume is no longer an invariant 
of the motion if shear viscosity is taken into account. 
Rather, the volume becomes a function of the emergent metric. 
This change of the volume corresponds 
 to dissipation in the real world. The equivalence 
between dissipation and introduction of external gravitational field 
seems apparent
 only at the first step, once we ``switch on'' viscosity. 
The gravitational field is a function of the
initial distribution of masses an their velocities. 
On the next step,  this adjustment should be reiterated. We have no 
proof that it is in fact  possible. Validity of our equation (\ref{crucial})
in higher orders in the effective gravitational field
is a guess made on physical grounds.  Note also
that generically dissipation results in appearance of imaginary parts 
of various correlators. Eq. (\ref{crucial}) does not involve
any imaginary parts. Physicswise, the reason is that conservation of charge
is not affected by dissipation. Loss of unitarity is manifested through
evolution of a real
quantity, that is physical volume.

In the next section, devoted to instabilities of 
chiral fluids we tried to make two points. 
First, we emphasised similarity of the dynamics 
in far infrared which might drive the chiral 
fluid to decay into helical magnetic and vortical configurations. 
Since the magnetic-field 
instability seems to be established theoretically, this similarity supports 
the idea on possible decay into vortical configurations as well.
Finally, we argued that the same similarity 
suggests that the emergent gravitational field
becomes a dynamical degree of freedom once the vortical instability is considered.

\section{Acknowlegements}

These notes were originally prepared for presentation
as a review talk at the conference
devoted to chiral dynamics at Tours, May 2016
organized by Le Studium foundation.  To a large extent, the notes are 
based on original papers written in collaboration with A. Avdoshkin, P. G. Minkin,
V. P. Kirilin and A. V. Sadofyev. The work was supported by Russian Science Foundation
Grant No 16-12-10059.


\begin{thebibliography}{99}
\bibitem{volume}
D. E. Kharzeev, K. Landsteiner, A. Schmitt, and H.-U. Yee,
``{\it Strongly Interacting Matter in Magnetic Field},
Lect. Notes Phys. 871, 1 (2013).


\bibitem{adler}
S. Adler, {\it ``Axial-Vector Vertex in Spinor Electrodynamics''}
Phys. Rev. 177 (1969) 2426;\\
J.S. Bell and R. Jackiw,{\it
"A PCAC puzzle: $\pi^0\to\gamma\gamma$ in the $\sigma$-model}". 
Nuovo Cim. A 60 (1969) 47. 


\bibitem{cheianov}
A. Yu. Alekseev, V. V. Cheianov, and J. Frohlich, 
{\it ``Universality of transport properties in equilibrium, 
Goldstone theorem and chiral anomaly''},
 Phys. Rev. Lett. 81 (1998) 3503, 
cond-mat/9803346. 



\bibitem{sonsurowka}
D.T. Son and P. Surowka,	
``{\it Hydrodynamcs 
with Triangle Anomalies}'',  Phys. Rev. Lett. 103 (2009) 191601, 
arXiv:0906.5044 [hep-th] .


\bibitem{oz}
Ya. Neiman and Ya. Oz, 
``{\it Relativistic Hydrodynamics with General Anomalous Charge}'',
  JHEP 1103 (2011) 023,
 arXiv:1011.5107 [hep-th].

\bibitem{erdmenger}
J. Erdmenger, M. Kaminski, and A. Yarom, 
``{Fluid dynamics of R-charged black holes},
 JHEP 0901 (2009) 055,  arXiv:0809.2488 [hep-th];\\
N. Banerjee, J. Bhattacharya, S. Bhattacharyya, S. Dutta, R. Loganayagam, and P. Surowka,
``{\it Hydrodynamics from charged black branes},
JHEP 1101 (2011) 094,
arXiv:0809.2596 [hep-th].
 
\bibitem{vilenkin}	
A. Vilenkin,
``{\it Equilibrium Parity Violating Current In A Magnetic Field},
 Phys. Rev. D22 (1980) 3080.


\bibitem{kharzeev1}
 D. E. Kharzeev, L. D. McLerran, and H. J. Warringa,  	
{\it ``The Effects of topological charge change in heavy ion collisions: 
'Event by event P and CP violation }'',
Nucl.
Phys. A 803 (2008) 227,  arXiv:0711.0950 [hep-ph];\\
K. Fukushima, D. E. Kharzeev, and H. J. Warringa, 
 {\it  	
``The Chiral Magnetic Effect''},
Phys. Rev. D 78 (2008) 074033,  arXiv:0808.3382 [hep-ph];\\
 D. Kharzeev, 
{\it  	
``Parity violation in hot QCD: Why it can happen, and how to look for it''}, 
Phys. Lett. B 633 (2006) 260, hep-ph/0406125.
 
\bibitem{vilenkin2}
A. Vilenkin,
{\it ``Quantum Field Theory At Finite Temperature In A Rotating System''},
Phys. Rev. D21 (1980) 2260.


 \bibitem{bekenstein}
H.K. Moffatt,
``{\it The degree of knottedness of tangled vortex lines}'', 
J. Fluid Mech. {\bf 35}, (1969) 117;\\
J. D. Bekenstein,
``{\it  Helicity conservation laws for fluids and plasmas}''
  Astrophys. Journ. {\bf 319} (1987) 207. 

\bibitem{golkar}
S. Golkar and D.T. Son,
``{\it (Non)-renormalization of the chiral vortical effect coefficient}'',
 JHEP 1502 (2015) 169,  arXiv:1207.5806 [hep-th].

\bibitem{nicolis} 	
S. Dubovsky, T. Gregoire, 
A. Nicolis, and R. Rattazzi, 
``{\it ``Null energy condition and superluminal propagation}'' ,
 JHEP 0603 (2006) 025,
hep-th/0512260;\\ 
S. Endlich, A. Nicolis , R. Rattazzi, and J. Wang, 
``{\it The quantum mechanics of perfect fluids}'',
JHEP, {\bf 1104}, 102 (2011),
arXiv:1011.6396 [hep-th];\\
S. Dubovsky, L. Hui, A. Nicolis, and D. T. Son,  
``{\it Effective field theory for hydrodynamics: thermodynamics, 
and the derivative expansion}'',
Phys. Rev. D85 (2012) 085029,
 arXiv:1107.0731 [hep-th].

 
  \bibitem{abanov}
G. M. Monteiro, A. G. Abanov, V.P. Nair,   
 	``{\it Hydrodynamics with gauge anomaly: Variational 
principle and Hamiltonian formulation}''
 Phys. Rev. D91 (2015), 125033,
 arXiv:1410.4833.
 

\bibitem{bhattacharaya}
S. Bhattacharyya, J. R. David, and S. Thakufr,
{\it `` Second order transport from anomalies''},
 JHEP 1401 (2014) 010,
  arXiv:1305.0340 [hep-th].


\bibitem{rangamani} 	
F. M. Haehl, R. Loganayagam, and M. Rangamani, 
{\it ``Effective actions for anomalous hydrodynamics''},
JHEP 1403 (2014) 034, 
arXiv:1312.0610 [hep-th].


\bibitem{avdoshkin}	
A. Avdoshkin, V.P. Kirilin, A.V. Sadofyev, and V.I. Zakharov, 
{\it ``On consistency of hydrodynamic approximation for chiral media''},
Phys. Lett. B755 (2016) 1,
 arXiv:1402.3587 [hep-th].

\bibitem{skenderis}
G. Compere, P. McFadden, K. Skenderis, and M. Taylor,
``{\it The Holographic fluid dual to vacuum Einstein gravity''},
JHEP 1107 (2011) 050,  arXiv:1103.3022 [hep-th].
 
\bibitem{shevchenko}
A.V. Sadofyev, V.I. Shevchenko, and V.I. Zakharov, 	
{\it ``Notes on chiral hydrodynamics within effective theory approach''},
Phys. Rev. D83 (2011) 105025, 
arXiv:1012.1958 [hep-th].

\bibitem{luttinger}
 J. M. Luttinger,
``{\it Theory of Thermal Transport Coefficients}'',
 Phys. Rev. 135A (1964) 1505.

\bibitem{stone}
 M. Stone, 
``{\it Gravitational Anomalies and Thermal Hall effect 
in Topological Insulators}'', Phys. Rev. B85 (2012) 184503,
 arXiv:1201.4095 [cond-mat.mes-hall].


\bibitem{zahed} 
I. Zahed, 	
{\it ``Anomalous Chiral Fermi Surface''},
Phys. Rev. Lett. 109 (2012) 091603,
 arXiv:1204.1955 ;\\
G. Basar, D. E. Kharzeev, and I. Zahed,
{\it ``Chiral and Gravitational Anomalies on Fermi Surfaces''},
 Phys. Rev. Lett. 111 (2013) 161601, 
arXiv:1307.2234.

\bibitem{liu}	
M. Crossley, P. Glorioso, and H. Liu,
``{\it Effective field theory of dissipative fluids}'',
arXiv:1511.03646 [hep-th] 

\bibitem{landau}
L. D. Landau and E. M. Lifshitz, ``Classical theory of Fields'', Pergamon.

\bibitem{alvarezgaume}
 L. Alvarez-Gaume  and E. Witten,	
``{\it Gravitational Anomalies}'',
 Nucl. Phys. B234 (1984) 269.

\bibitem{minkin}
P. G. Minkin and V.I.  Zakharov, in preparation.

\bibitem{goldstone} 	
J. Goldstone and F. Wilczek,
``{\it Fractional Quantum Numbers on Solitons}'',
 Phys. Rev. Lett. 47 (1981) 986.

\bibitem{harvey}	
C. G. Callan, Jr. and  J. A. Harvey,
``{\it Anomalies and Fermion Zero Modes on Strings and Domain Walls}'',
 Nucl. Phys. B250 (1985) 427.

\bibitem{wilczek}
S. P. Robinson and F. Wilczek,  
 	``{\it A Relationship between Hawking radiation and gravitational anomalies}'',
 Phys. Rev. Lett. 95 (2005) 011303

\bibitem{redlich}
A.N. Redlich, 
 ``{\it Gauge Noninvariance and Parity Violation of Three-Dimensional Fermions}'',
 Phys. Rev. Lett. 52 (1984) 18;\\
 Yu. Akamatsu and N. Yamamoto, 
``{ \it Chiral Plasma Instabilities}'',
 Phys. Rev. Lett. 111 (2013) 052002, 
arXiv:1302.2125 [nucl-th]. 

\bibitem{khaidukov}
Z.V. Khaidukov, V.P. Kirilin, A.V. Sadofyev, and V.I. Zakharov, 
``{\it On Magnetostatics of Chiral Media}'',
arXiv:1307.0138 [hep-th]. 

\bibitem{yamamoto} 
N. Yamamoto, ``{\it 	
Chiral transport of neutrinos in supernovae: Neutrino-induced fluid helicity and 
helical plasma instability}'',
Phys. Rev. D93 (2016)  065017,
arXiv:1511.00933 [astro-ph.HE].

 

\bibitem{burch}	
T. Burch and  G. Torrieri, 
``{ \it Indications of
  a non-trivial vacuum in the effective theory of perfect fluids}'',
 Phys. Rev. D92 (2015) 1, 016009, 
arXiv:1502.05421 [hep-lat].

\bibitem{kirilin1}
V.P.Kirilin, A.V. Sadofyev and V.I. Zakharov, in preparation. 

\bibitem{dvali}	
G. Dvali,  G. F. Giudice, C. Gomez, and A. Kehagias,
``{\it UV-Completion by Classicalization}'',
JHEP 1108 (2011) 108, 
arXiv:1010.1415 [hep-ph].


\bibitem{page}
S.W. Hawking and D. N. Page, 
``{\it Thermodynamics of Black Holes in anti-De Sitter Space}'',
Commun. Math. Phys. 87 (1983) 577. 

\end{thebibliography}
\end{document}